\documentstyle[prl,multicol,twoside,epsf,epsfig,aps]{revtex}
\title{Strong friction limit in quantum mechanics: the Quantum
Smoluchowski Equation}  

\author{Joachim Ankerhold, Philip Pechukas$^*$, and Hermann Grabert}
\address{
Fakult{\"a}t f{\"u}r Physik, Albert-Ludwigs-Universit{\"a}t
Freiburg,
 Hermann-Herder-Stra{\ss}e 3, D-79104 Freiburg, Germany}

\date{Phys.~Rev.~Lett., in press}

\begin{document}

\maketitle

\begin{abstract}
For a quantum system coupled to a heat bath environment the strong
friction limit is studied starting from the exact path integral
formulation. Generalizing the classical Smoluchowski limit to low
temperatures a time
evolution equation for the position distribution is derived and the
strong role of quantum fluctuations in this limit is revealed. 
\end{abstract}

\vspace{-0.3cm}
\hspace{1cm}PACS:03.65.Yz,05.30.Ch,73.23.-b,82.20.-w

%\vspace*{-0.5cm}
\begin{multicols}{2}
Quantum systems coupled to a heat bath environment can be found almost
everywhere in physics and chemistry \cite{weiss}. Transport processes in
Josephson junctions \cite{kramers}  or electron transfer reactions in
large molecules \cite{marcus}
are typical examples. What one aims to describe here is the effective
dynamics of the relevant system degrees of freedom, i.e.\ the reduced
dynamics.
While the corresponding classical theory is well established and based
on Fokker--Planck equations, the formulation of dissipation 
 in quantum mechanics is more complicated. In general a
simple time evolution equation for the reduced density 
matrix does not exist\cite{weiss}; 
a formally exact expression for the reduced dynamics 
 in terms of path integrals is available, but the path integral
expression in many cases cannot be evaluated even numerically.
In the last years efforts have focused on the weak coupling regime
where a description 
in terms of Master equations is possible \cite{master}.
The opposite limit
of strong coupling has been left basically untouched \cite{annalen}. 
Here we
study this limit. For the first time we show that for large friction
the exact dynamics of a dissipative 
quantum system can be  cast into a time evolution equation for
the  position distribution, the so-called quantum
Smoluchowski equation derived below in Eq.~(\ref{qmsmolu}).
 
In classical physics the Smoluchowski limit is well-known
in all areas  \cite{risken}. For large friction the
 Fokker-Planck equation 
in phase space reduces to a time evolution equation in position space,
the Smoluchowski equation (SE). There, the basic condition is a time scale
separation between relaxation of momentum and position which allows
for an adiabatic elimination of the former degree of
freedom. Corrections to the Smoluchowski equation turn out to be
algebraically small in the friction 
constant. Applications are countless; recent examples include
transport in systems with fluctuating barriers \cite{fluc1} or
 ratchets \cite{magnasco} as found in tunnel diodes \cite{fluc2} and complex 
macromolecules \cite{astumian}, and decay in
 periodically driven metastable 
potentials \cite{osci}.  The quantum Smoluchowski equation  should be
 important for the description of similar processes at low
 temperatures. 

{\it Quantum dissipation--}The standard way to describe dissipation
in quantum mechanics is  
based on system+reservoir-models with a total Hamiltonian
$H=H_S+H_R+H_{I}$ \cite{weiss}. The reservoir $H_R$ (heat bath) consists of
a quasi-continuum of harmonic oscillators which are  coupled bilinearly
with the system $H_S$ 
via the interaction $H_I$. Classically,
this model leads back to a generalized Langevin
equation for the reduced  system. The 
quantum dynamics 
for the reduced density matrix follows from
 $\rho(t) = {\rm Tr}_R\{
\exp(-iHt/\hbar)$$W(0)\exp(iHt/\hbar)\}$ with the correlated initial
state $W(0)$. In the position representation the path integral
approach allows for an exact elimination of the bath 
degrees of freedom. Now, as in the classical Smoluchowski
limit we focus on the position probability distribution
$P(q,t)=\rho(q,q,t)$ and obtain the exact result \cite{weiss,report}  
\begin{equation}
P(q_f,t)=\int dq_i dq_i'\, 
J_t(q_f,q_i,q_i')\,  \Lambda(q_i,q_i').\label{density}
\end{equation}
Here, the propagating function $J_t$ is a three-fold path
integral (two in real, one in imaginary time) over the system degrees
of freedom only. The real time paths $q(s)$  and $q'(s)$ run in time
$t$ from
$q_i$ and $q_i'$  to the fixed endpoint $q_f$,
while the imaginary time paths
$\bar{q}(\sigma)$ connect $q_i$ with $q_i'$ in the 
 interval $\hbar\beta$ ($\beta=1/k_{\rm B} T$). 
The contribution of each path is weighted by
 $\exp(i\Sigma[\bar{q},q,q']/\hbar)$ with an effective action
$\Sigma$ \cite{weiss,report} not written down here explicitly. The
real time paths describe the dynamics of the  
system and the imaginary time paths specify the initial probability
distribution 
\begin{equation}
P(q_i,0)= \left[\,\rho_\beta(q_i,q_i')\, \Lambda(q_i,q_i')\,
\right]_{q_i=q_i'}\, ,\label{anfang} 
\end{equation} 
where
$\rho_\beta(q,q')=\langle q|{\rm Tr_R}\exp(-\beta
H)|q'\rangle$ is the reduced equilibrium density matrix while the
preparation function $\Lambda$ describes the deviations from
equilibrium. This way we avoid a factorizing initial state 
used in ordinary Feynman-Vernon theory \cite{weiss}, which does not apply for
strong friction. 
In Eq.~(\ref{density}) the influence of the bath 
 is completely determined by the spectral density $I(\omega)$
of the bath oscillators. Effectively, the reduced system
gains an additional interaction contribution (influence functional)
which is non-local in time and strongly depends on friction and
temperature. Since the influence functional couples real and
imaginary time paths,  
an  exact solution to 
Eq.~(\ref{density})  is accessible only in certain
cases. For weak damping the Born-Markov
approximation can be invoked leading to the known Master
equations  \cite{master}.
Here, we explore to what extent the dynamics in
(\ref{density}) can 
be well approximated by a time evolution equation when the friction is
large. For a particle of mass $M$ we assume Ohmic
damping $I(\omega)= 2 M \gamma$ (damping constant $\gamma$) supplemented  by a
high frequency cut-off $\omega_c>\gamma$ according to spectral
densities in real systems and to avoid the well-known 
ultraviolet divergencies of a pure Ohmic model. Further, as in the
classical SE we take sufficiently smooth potentials for granted in this
study.

{\it Harmonic oscillator--} For the
quantum Brownian motion in
the harmonic potential $V(q)=\frac{1}{2} 
M\omega_0^2 q^2$
 the propagating function is known exactly 
\cite{weiss,report}. To get insight in the Smoluchowski simplification
we particularly study the relaxation of $\langle q(t)\rangle$ and
$\langle p(t)\rangle$ in the overdamped limit  $\gamma/\omega_0\gg 1$ and for
times $t\gg 1/\gamma$. One finds $\langle q(t)\rangle \approx
\langle q(t)\rangle_{\rm cl}+\delta q(t)$ and
$\langle p(t)\rangle \approx \langle q(t)\rangle_{\rm cl}/\gamma+M\delta
\dot{q}(t)$. The classical part shows the well-known sluggish
decay $\langle q(t)\rangle_{\rm cl}\propto \exp(-\omega_0^2
t/\gamma)$ and to leading order $\langle p(t)\rangle\approx
M\delta\dot{q}(t)$. Quantum effects show up in 
 $\delta q(t)$ which depends on the
initial correlations between system and bath and for
temperatures $T>0$ decays 
as $\exp(-\nu t)$ where $\nu=2\pi/\hbar\beta$ is the Matsubara
frequency. Hence, 
on the time scale where the position relaxes, i.e.\
$\gamma/\omega_0^2$,  
the momentum can be considered as already equilibrated, i.e.\ $\langle
p(t)\rangle\approx 0$,  only if a separation of time scales is guaranteed,
i.e.\  $\gamma/\omega_0^2\gg \hbar\beta, 1/\gamma$ [cf.\
Fig.~1]. This latter condition 
defines the Smoluchowski range also for temperatures where
quantum fluctuations are important. Now, neglecting 
terms of order $\exp(-\gamma t)$ and $\exp(-\nu t)$ or
smaller the 
 time evolution for $P(q,t)$ follows from
\begin{equation}
\partial_t
P(q,t)=({\omega_0^2/\gamma}){\partial_q}\left[q+\langle
q^2\rangle{\partial_q}\right] P(q,t)\label{harmsmolu}
\end{equation}
where the variance reads
\begin{equation}
\langle q^2\rangle \approx \frac{1}{M\omega_0^2 \beta}+\frac{2}{M\beta}
\sum_{n\geq 1} \frac{1}{\nu_n^2+\nu_n \gamma},\ \ \nu_n=n\nu.\label{q2}
\end{equation}
Inspection of the sum in Eq.~(\ref{q2}) reveals that 
 the Smoluchowski range
$\gamma/\omega_0^2\gg \hbar\beta, 1/\gamma$ comprises two particular
 subsets: one is the known classical range, $\gamma\ll
\nu$, the other is a  quantum mechanical region
with $\gamma\gg \nu$, unstudied so far (Fig.~1). While in both
 ranges to leading 
order $\langle q^2\rangle\approx\langle q^2\rangle_{\rm cl}$,  
differences appear in the respective  corrections. Classically, these
 are of order $\hbar^2\beta/M$ and negligible against
 dynamical corrections of order $1/M\beta\gamma^2$. In 
contrast, in the quantum range to first order
\begin{equation}
\langle q^2\rangle-\langle q^2\rangle_{\rm cl}\approx
({\hbar}/{\pi
M\gamma})\, \log(\hbar\beta\gamma/2\pi)\equiv\lambda\label{lambda}
\end{equation}
with $\lambda\gg 1/M\beta\gamma^2$. We conclude that for
$\hbar\gamma\gg k_{\rm B} T$ the dynamics of the quantum 
oscillator follows in leading order from the classical SE
with quantum fluctuations as the dominating corrections.
Since these  are no longer
algebraically small, they are of substantial relevance (see below). We
note in passing that in the quantum range one  
finds $\langle 
p^2\rangle \approx (M\hbar\gamma/\pi)[1+\log(\omega_c/\gamma)]$.
\vspace*{-2cm}
\begin{figure}
%\vspace{-1cm}
\center
\epsfig{file=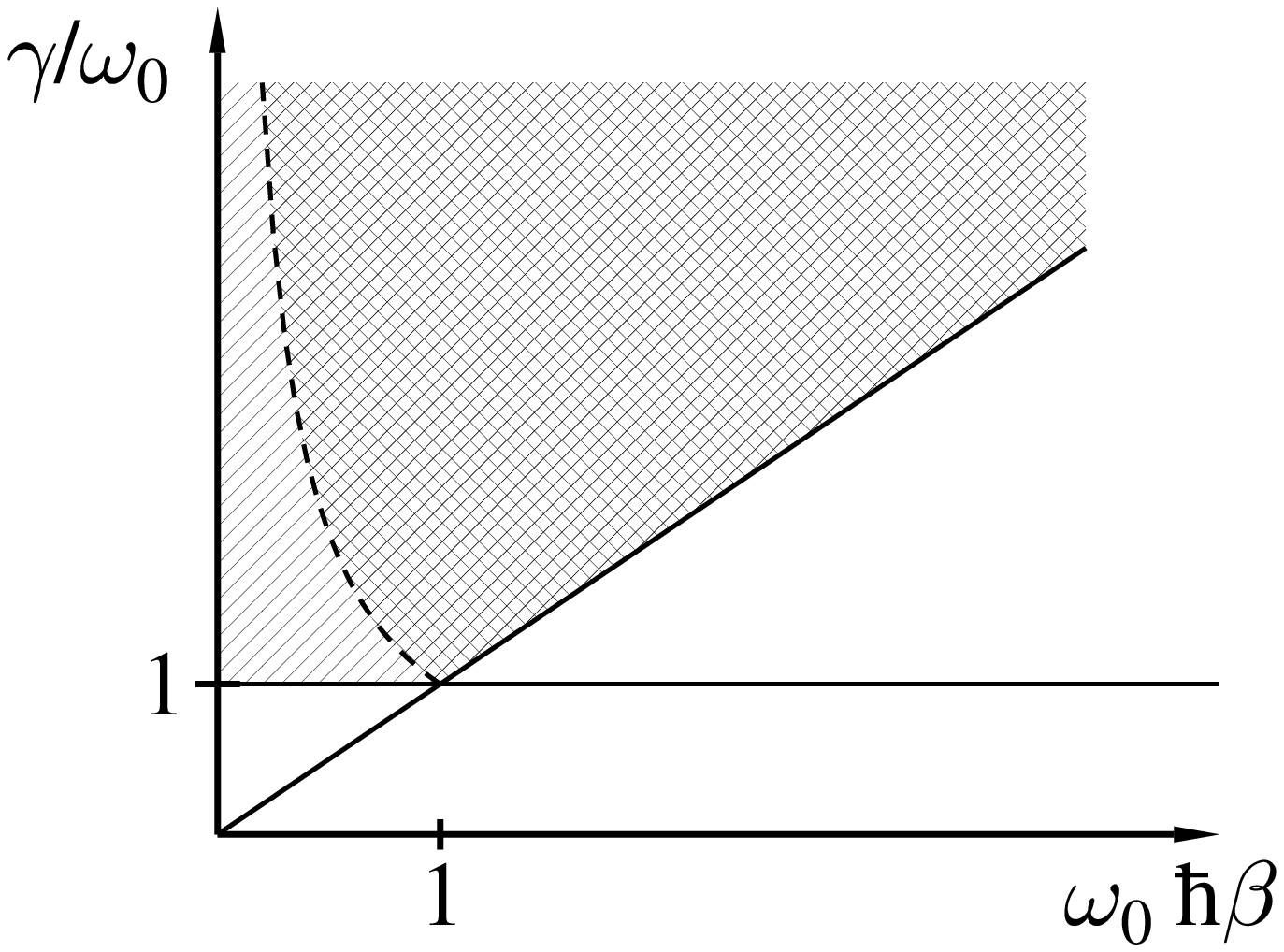,width=8cm}
\vspace{-5.2cm}
%\parbox{5cm}{\caption{Smoluchowski range $\gamma/\omega_0^2\gg
%\hbar\beta,1/\gamma$ (shaded). The classical range
%($\gamma\ll\nu$) is simple shaded, the quantum range ($\gamma\gg\nu$)
%double shaded.}
\label{fig1}
\end{figure}
\noindent
\parbox{8.5cm}{\small{FIG.~1 Smoluchowski range $\gamma/\omega_0^2\gg
\hbar\beta,1/\gamma$ (shaded). The classical range
($\gamma\ll\nu$) is simple shaded, the quantum range ($\gamma\gg\nu$)
double shaded.}}

\vspace*{0.25cm}
{\it Quantum Smoluchowski equation--}Qualitatively, to leading order 
we expect the position diagonal element $P(q_f,t)$ to obey  the
classical SE even in anharmonic potentials and throughout the quantum
Smoluchowski range (QSR) 
\begin{equation}
 \gamma/\omega_0^2\gg\hbar\beta, 1/\gamma\ \ \ \ \mbox{ and} \ \ \ \
\hbar\gamma\gg 
k_{\rm B} T,\label{aq1}
\end{equation}
 where $\omega_0$ is interpreted as
the ground state frequency.  
To quantify this
statement we analyze  
 the path integrals involved in Eq.~(\ref{density}) in more
detail. For that purpose, it 
is convenient to introduce difference and sum real-time paths $x=q-q'$
and $r=(q+q')/2$, respectively. Then, it turns out that the action
$\Sigma[\bar{q},x,r]$ has friction dependent parts in
imaginary and in real time \cite{report} which suppress non-diagonal matrix
elements. Accordingly, to leading order we assume
$|\bar{q}(\sigma)-\bar{q}(0)|$ and $|x(s)|$ to be sufficiently small and
later verify this ansatz self-consistently.

First, by expanding
$\Sigma[\bar{q},x,r]$ up to second order in the 
$x$-paths, we arrive at a solvable Gaussian $x$-path integral where
$x$ and $r$ paths are coupled bilinearly.
We put $x(s)=x_i+\delta x(s)$ with fluctuations
$\delta x(s)$ obeying $\delta x(0)=\delta x(t)=0$ and to leading order
obtain 
\begin{equation}
\int {\cal D}x {\rm e}^{i(\Sigma[\bar{q},x,r]-
\Sigma[\bar{q},0,0])/\hbar}  \approx \delta(x_i){\rm
e}^{-S[r]/4M\gamma k_{\rm B} T}\label{aq2}
\end{equation}
with the action 
\begin{equation}
S[r]=\int_0^t ds \left[M \gamma\dot{r}+V'(r)\right]^2.\label{action}
\end{equation}
The path integral over the imaginary time path $\bar{q}$ with
$\bar{q}(0)=\bar{q}(\hbar\beta)=r_i$ is easily evaluated. It
yields  a contribution 
proportional to $\exp[-\beta V(r_i)]$.
  Eventually,
Eq.~(\ref{density}) for $P(q_f,t)=\rho(q_f,q_f,t)$ reduces to
\begin{equation}
P(q_f,t)=\int dq_i\ G(q_f,t,q_i)\, P(q_i,0).\label{leadingsm}
\end{equation}
Here, the propagator stems from the remaining $r$-path integral
and thus,  is a sum over all paths $r(s)$
with $r(0)=q_i, r(t)=q_f$  where each contribution is weighted with
$\exp(-S[r]/4M\gamma k_{\rm B} T)$. 
However, this is exactly  the path integral representation of
the classical Smoluchowski dynamics \cite{risken}.
 Hence, to leading order the
quantum SE looks like the classical one also for
anharmonic potentials: $\partial_t
P=(1/M\gamma)\partial_q \hat{L}_{\rm 
cl}P$ with
\begin{equation}
\hat{L}_{\rm cl}=V'+k_{\rm B}T{\partial_q}\label{clsmolu} 
\end{equation} 
where $V'(q)=d V(q)/dq$. To justify our initial assumption we estimate that
$[\bar{q}(\sigma)-\bar{q}(0)]^2$ is at most of order $\lambda$ [see
Eq.~(\ref{lambda})] and $x(s)^2$ at most of order
$\hbar^2\beta/M\gamma t$.

Of course,  the most interesting issue
is to specify contributions from  quantum fluctuations to Eq.~(\ref{clsmolu}).
However, before we do so let us first look at the simpler static problem 
and determine the equilibrium distribution.
In the relevant parameter range the path integral for
$P_\beta(q)=\rho_\beta({q},{q})$ is calculated by putting
$\bar{q}(\sigma)=\bar{q}_{\rm ma}(\sigma)+y(\sigma)$.
The minimal action
path  $\bar{q}_{\rm ma}$ with $\bar{q}_{\rm ma}(0)=\bar{q}_{\rm
ma}(\hbar\beta)=q$  and its Euclidian action $S[\bar{q}_{\rm
ma}]$ are evaluated up terms linear in $\lambda$.
The result is $S(q)/\hbar=\beta[V(q)-\lambda\beta V'(q)^2/2]$. The
path integral over $y$ is approximated by a Gaussian 
integral since $(\bar{q}_{\rm ma}-y)^2{\textstyle
 {\lower 2pt \hbox{$<$} \atop \raise 1pt \hbox{$ \sim$}}}O(\lambda)
$ and is solved by expanding $y(\sigma)=\frac{1}{\hbar\beta}\sum y_n
\exp(i\nu_n \sigma)$ with the boundary
condition $\Sigma y_n=0$.  As an important result we obtain the 
unnormalized equilibrium position distribution in the QSR
\begin{equation}
P_\beta(q)= P_\beta^{\rm cl}(q)\ \exp\left\{\lambda\beta\left[\beta
{V'(q)}^2/2-3 V''(q)/2\right]\right\}\label{equi}
\end{equation}
with $P_\beta^{\rm cl}(q)=\exp[-\beta V(q)]$. 
 Next order corrections are much smaller, namely, of order $
 (\hbar\beta V''/M\gamma)^2$. Amazingly, for low temperatures $P_\beta$ will
 be squeezed by friction without limit.

Let us now turn to the dynamical aspect of the QSR and evaluate
quantum fluctuations to Eq.~(\ref{clsmolu}). For this purpose we write
$\partial_t P=(1/M\gamma)\partial_q \hat{L}_{\rm qm} P$ with
$\hat{L}_{\rm qm}=\hat{L}_{\rm cl}+\lambda\delta \hat{L}$ where
$\hat{L}_{\rm cl}$  is given in (\ref{clsmolu})  and $\delta
\hat{L}=\delta V'+\delta D\partial_q$ with
appropriate corrections to the potential and diffusion terms.
To determine $\delta V$ and $\delta D$ 
requires to solve the full
dynamical problem 
Eq.~(\ref{density}) perturbatively in $\lambda$. Further, we need only
to study
the broad time range $\hbar\beta, 1/\gamma\ll t<\gamma/\omega_0^2$.
Once we have $\delta \hat{L}$ in this 
time interval, we have it for all times as the time scale
separation renders the operator $\hat{L}_{\rm qm}$ 
time independent on the coarse grained time scale (\ref{aq1}).
The general strategy is as above: evaluate the
minimal action paths to $\Sigma[\bar{q},x,r]$ and the corresponding
fluctuations  including corrections of order $\lambda$. Again, 
a small $x$ expansion of $\Sigma$  allows us to neglect
higher than bilinear couplings between $x$ and $r$-paths.
Within the relevant time range it is also consistent to assume the
difference $|r(s)-r_i|$ to be sufficiently small so that we may write
 $V[r(s)]\approx
V(r_i)+V'(r_i)[r(s)-r_i]+V''(r_i)/2[r(s)-r_i]^2$. 

 The solution to
the classical $x$-path is then simply $x_{\rm cl}(s)=x_i
\exp[-s V''(r_i)/M\gamma]$. Since the  $r$-path is coupled to both
the $x$ and the $\bar{q}$ path, the expression for $r_{\rm cl}$ is
rather lengthy and not 
given here. The same holds true for the imaginary time path, which can
be put in the form
$\bar{q}(\sigma)=r_i+\lambda \bar{q}_1[\sigma;x]$ where $\bar{q}_1$
 contains also the coupling to the $x$-path. As for the equilibrium case,  
the fluctuation path integrals can be evaluated in a  Gaussian
approximation: The contribution from the imaginary time path reads as
 in Eq.~(\ref{equi}), while the real time path integrals
provide a factor of order $\gamma$. 
As a result we gain the propagating function in the
QSR which determines $P(q,t)$ via
ordinary integrations over the initial coordinates $x_i$ and $r_i$, see 
Eq.~(\ref{density}). Now,
further progress can be made by introducing the scaled coordinate
$k_i=x_i\gamma$.  Then, as a function of $k_i$ the propagating
function is basically a Gaussian with a width of order
$M\gamma\beta/t$, so that for sufficiently smooth
preparations in $x_i$-space, i.e.\ sufficiently localized preparations
in momentum space, it is consistent to set
${\Lambda}(k_i/\gamma,r_i)\approx {\Lambda}(0,r_i)$.
This allows us to carry out  the $k_i$ integration. The time evolution
of $P(q,t)$ 
 in the QSR and within the relevant time
range is therefore of the form (\ref{leadingsm}) with
the propagator  
\begin{equation}
G(q,y,t)= \frac{1}{\sqrt{4\pi\sigma(y,t)}}\ {\rm
e}^{-[q-r(y,t)]^2/4\sigma(y,t)}.\label{aq4} 
\end{equation} 
Here, the  width is
$\sigma=\sigma_0+\lambda\sigma_1$ with 
$\sigma_0=(t/M\gamma\beta)[1-V'' t/M\gamma]$ and $\sigma_1=-[1-V''
t/M\gamma]$, and the local motion reads $r=r_0+\lambda r_1$ with $r_0=y-[V'
t/M\gamma](1-V''t/2M\gamma)$ and $r_1=y \beta  V'' \sigma_1$.

The propagator (\ref{aq4}) is now used to determine  $\delta V'$ and
$\delta D$ 
in $\hat{L}_{\rm qm}$. Starting from Eq.~(\ref{leadingsm}) one applies
$\partial_t-(1/M\gamma)\partial_q \hat{L}_{\rm qm}$ to $G(q,y,t)$ and then
expands $y$ dependent terms in the integral up to second 
order around $y=q$. After tedious algebra we find
 $\delta V'=V'''/2$ and $\delta D=\beta V''$. This leads us
to the central result of this paper, namely, the dynamical
equation of an 
overdamped quantum system, the so-called
quantum Smoluchowski equation (QSE), $\partial_t P=(1/M\gamma)\partial_q
\hat{L}_{\rm qm} P$ with
\begin{equation}
\hat{L}_{\rm qm}=V'+\lambda V'''/2+\partial_q\, k_{\rm
B}T[1+\lambda\beta V'']. \label{qmsmolu}
\end{equation}
Two observations can be made here: First, 
 quantum effects give rise to an
effective potential
$V_{\rm eff}=V+\lambda V''/2$ and an effective diffusion term $D_{\rm
eff}=k_{\rm B}T[1+\lambda\beta V'']$. Second,  seen as a continuity equation
it turns out that the equilibrium distribution
Eq.~(\ref{equi})   indeed fulfills $\hat{L}_{\rm
 qm}P_\beta=0$. Corrections to the QSE are of order 
 $\beta (V''\hbar)^2/M^3\gamma^3 $ and can thus savely 
be disregarded.

Now, from the QSE the
quantum analog to the classical  Langevin equation is (in the
Stratonovich sense \cite{hangg})  
\begin{equation}
M\gamma \dot{q}(t)+V'(q)+\lambda V'''=\sqrt{1+\lambda\beta
V''}\ \xi(t)\label{cllangevin} 
\end{equation}
with Gaussian noise $\langle
\xi(t)\rangle=0$, 
$\langle \xi(t) \xi(t')\rangle=(2M\gamma/\beta)\delta(t-t')$. 
Hence, the quantum stochastic
process in the QSR is 
equivalent to a classical process in an effective potential and with
multiplicative noise.  
 
{\it Applications--}We illustrate the 
significance of quantum fluctuations in the QSR in two paradigmatic
applications. 
 
Imagine a metastable barrier 
potential $V(q)$ where the barrier height $V_b$ is 
 the largest energy scale in the system, i.e.\ $V_b\gg k_{\rm B}
T$. Then, preparing the 
system initially in an equilibrium state restricted to the well
region (around $q=0$),  for intermediate times (plateau
range) the distribution becomes quasi-stationary $P(q,t)\to P_{\rm
st}(q)$. This state corresponds to a constant flux across the barrier
(around $q=q_b$) $
J_{\rm st}=(1/M\gamma)\hat{L}_{\rm qm} P_{\rm
st}$ determining the escape rate through $\Gamma=J_{\rm st}/N$ with
the normalization $N$ given by the well population. 
Now, adopting the classical procedure we obtain from Eq.~(\ref{qmsmolu})
\begin{equation}
\Gamma_{QSR}= \frac{\sqrt{V''(0)|V''(q_b)|}}{M\gamma} {\rm
e}^{-\beta V_b}\, {\rm e}^{\lambda\beta
[V''(0)+|V''(q_b)|]}\label{qserate}
\end{equation}
Here,  the second factor accounts for
the quantum 
fluctuations and is not small at all (see Fig.~2). The point
is that $\lambda$ dependent terms 
 in the QSE  enter exponentially and thus lead to a
substantial rate increase compared to the
classical rate. Already for moderate friction $\Gamma_{QSR}$ agrees
well with the
 exact rate (for $V_b/k_{\rm B} T\gg 1$) \cite{wolynes}. 
\vspace*{-1cm}
\begin{figure}
\center
\epsfig{file=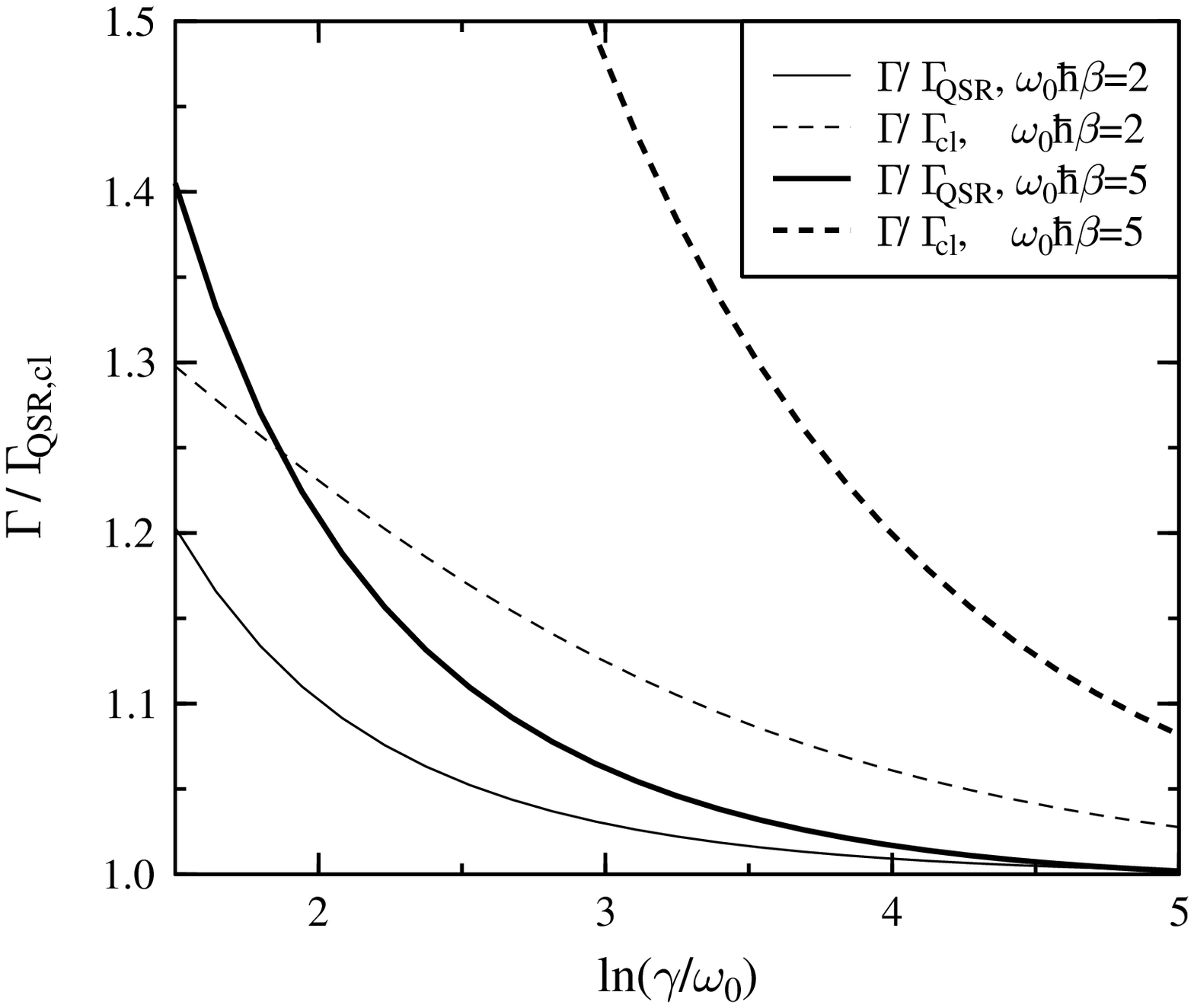,width=7.5cm}
\vspace{-4.4cm}
%\caption{Ratio of the exact rate with the
%classical (dashed) and the Smoluchowski (solid) rate vs. friction for
%$|V''(q_b)|=V''(0)=M\omega_0^2$.
\label{fig2}
\end{figure}
\noindent
\parbox{8.5cm}{\small{FIG.~2 Ratio of the exact rate with the
classical (dashed) and the Smoluchowski (solid) rate vs. friction for
$|V''(q_b)|=V''(0)=M\omega_0^2$.}}
 
\vspace*{0.25cm}
In order to analyze  $V_{\rm eff}$ and
$D_{\rm eff}$ in more detail we consider the
dynamics of a Gaussian wave packet in a bistable potential
$V(q)=-M\omega_0^2 q^2/2[1-q^2/2q_a^2]$. The barrier top is located at
$q=0$ and the wells at $q=\pm q_a$.
%,and the barrier height is
%$V_b=M\omega_0^2 q_a^2/4$. 
Then, scaling times with
$\gamma/\omega_0^2$ and coordinates with $\sqrt{\hbar/M\omega_0}$ we solve
the time evolution of a wave packet initially localized near the barrier top
 numerically also for longer times. High precision results are
presented in Fig.~3 for $\langle q(t)\rangle$ at
two different temperatures. 
%and fixed small $\lambda=0.035$. 
For moderate temperature
 the influence of $V_{\rm eff}$
prevails: the barrier height is diminished, i.e.\ $V_{{\rm
eff},b}=V_b-3\lambda M\omega_0^2/2$, thus increasing the
population around $q=0$ so that $\langle q(t)\rangle <\langle
q(t)\rangle_{\rm cl}$. In contrast, at low temperatures diffusion
effects dominate as the
 spreading of the wave packet near $q=0$ is retarded since
the local effective  temperature  
%$T_{\rm eff}(0)=T[1-\lambda\beta |V''(0)|]$
is lowered $T_{\rm eff}(0)<T$, i.e.\  
$\langle q(t)\rangle >\langle
q(t)\rangle_{\rm cl}$. 
\vspace*{0cm}
\begin{figure}
\center
\epsfig{file=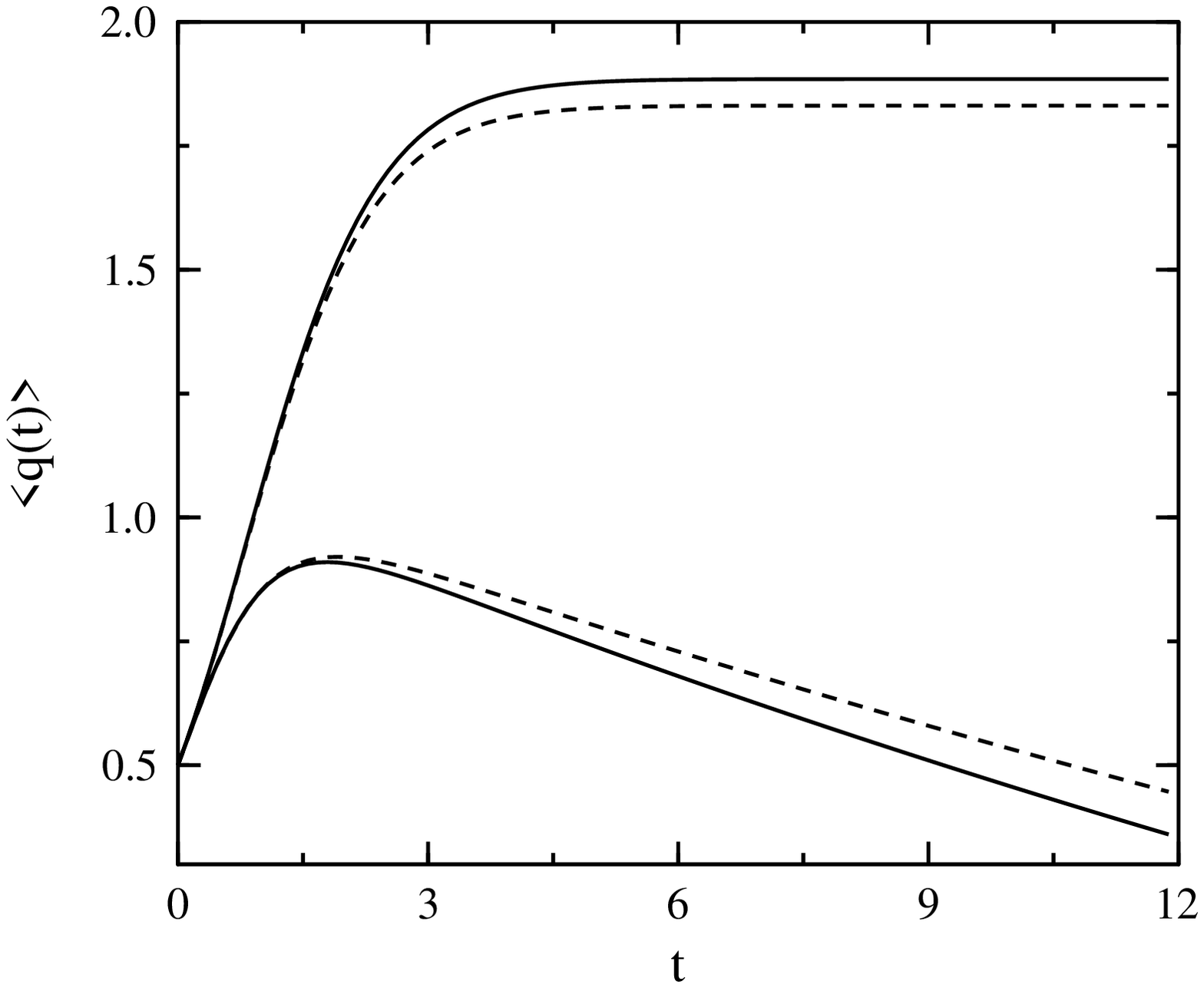,width=7.25cm}
\vspace{-4.45cm}
%\caption{FIG.~3 Mean position vs. time of a  Gaussian wave packet with
%$\langle q(0)\rangle=0.5$ moving in a bistable 
%potential. Dashed (solid) lines are classical (quantum) 
%results with $\lambda=0.035$. Lower (upper) curves are for
%$\omega_0\hbar\beta=1$ ($\omega_0\hbar\beta=6$).
\label{fig3}
\end{figure}
\noindent
\parbox{8.5cm}{\small{Mean position vs. time of a  Gaussian wave packet with
$\langle q(0)\rangle=0.5$ moving in a bistable 
potential. Dashed (solid) lines are classical (quantum) 
results with $\lambda=0.035$. Lower (upper) curves are for
$\omega_0\hbar\beta=1$ ($\omega_0\hbar\beta=6$).}}

\vspace*{0.25cm}
To summarize we have derived the time evolution
equation for dissipative quantum systems in the strong friction regime
and have shown the substantial impact of quantum fluctuations.
As the classical Smoluchowski limit is of importance for all
overdamped systems at high temperatures, our results  
have  potentially an equally wide range of applications for overdamped
quantum systems in low temperature physics and chemistry.

This work was supported by the DFG (Bonn) through SFB276 and by the
National Science Foundation, under grants CHE-0078632 and INT-9726203.

\end{multicols}
\end{document}